\documentclass[letterpaper]{article} 
\usepackage{aaai2026}  
\usepackage{times}  
\usepackage{helvet}  
\usepackage{courier}  
\usepackage[hyphens]{url}  
\usepackage{graphicx} 
\usepackage{natbib}  
\usepackage{caption} 
\usepackage{algorithm}
\usepackage{algorithmic}

\usepackage{newfloat}
\usepackage{listings}
\DeclareCaptionStyle{ruled}{labelfont=normalfont,labelsep=colon,strut=off} 
\floatstyle{ruled}
\newfloat{listing}{tb}{lst}{}
\floatname{listing}{Listing}

\usepackage{array}
\usepackage{booktabs}
\newcolumntype{P}[1]{>{\centering\arraybackslash}p{#1}}
\usepackage{microtype}   
\usepackage{xcolor}     
\usepackage{amsmath}
\usepackage{makecell}
\usepackage{graphicx}
\usepackage{tikz}
\usetikzlibrary{positioning,arrows.meta}
\usepackage{pifont}
\usepackage{fontawesome5}
\usepackage{multirow}

\title{Grounded but Misleading: Evaluating Semantic Alignment in AI-Generated Security Explanations}
\author{
    Heajun An\textsuperscript{\rm 1},
    Connor Ng\textsuperscript{\rm 1},
    Sandesh Sharma Dulal\textsuperscript{\rm 1},
    Junghwan Kim\textsuperscript{\rm 1},
    Jin-Hee Cho\textsuperscript{\rm 1}
}

\affiliations{
    \textsuperscript{\rm 1}Virginia Tech\\
    \{heajun, connorng, sandesh339, junghwankim, jicho\}@vt.edu
}

\usepackage{bibentry}

\begin{document}

\maketitle

\begin{abstract}
Online scams increasingly leverage fluent and context-aware social-engineering strategies, creating growing demand for AI systems that explain why a message may be risky. However, explanations that cite detector-derived evidence may still semantically weaken or redirect the intended risk interpretation. We introduce \textbf{VEXA} (\textbf{V}erifying Semantic \textbf{EX}planation \textbf{A}lignment), a controlled testbed for studying the gap between lexical grounding and semantic risk alignment in AI-generated scam-risk explanations. VEXA generates ungrounded, risk-aligned, and risk-diluting explanations by independently controlling evidence grounding and semantic framing. Through LLM-as-a-judge and human evaluations, we show that explanations may continue to appear comparatively grounded even when their semantic interpretation weakens the detector’s intended risk assessment. In human evaluation, risk-diluting XAI-grounded explanations retained comparatively elevated Perceived Evidence Grounding scores ($3.66 \pm 1.02$) despite lower Helpfulness ($3.00 \pm 1.41$) and Reasoning Support ($3.14 \pm 1.05$) scores. These findings provide controlled evidence of grounding-illusion effects in AI-generated security explanations and suggest that trustworthy explanation evaluation must verify not only whether evidence is cited, but also how that evidence is interpreted.
\end{abstract}


\section{Introduction}
\label{sec:introduction}

Online scams have become a pervasive security challenge across emails, SMSs, and social media, causing billions of dollars in annual losses~\cite{FTC2025FraudData}. This challenge is further intensified by generative AI, which enables highly fluent and context-aware scam messages at scale~\cite{gressel2024discussion}. As AI-assisted scam detection systems become increasingly deployed, explanations are expected not only to provide binary warnings, but also to justify why a message may be risky.

\textbf{Explainable Scam Detection.}
Recent transformer-based detectors have substantially improved scam and phishing detection~\cite{uddin2024scam}. In parallel, eXplainable AI (XAI) methods such as SHAP~\cite{Lundberg2017Shap} and LIME~\cite{ribeiro2016should} identify influential scam-related cues. However, they remain largely token-level and model-centric, providing limited explanation for why such cues indicate deception or how users should interpret them in context~\cite{miller2019explanation}.

\textbf{From Lexical Grounding to Semantic Alignment.}
Large Language Models (LLMs) can transform detector-derived evidence into natural-language explanations. However, evidence citation alone does not guarantee detector-aligned interpretation. Even when explanations reference suspicious cues, they may weaken or redirect the detector’s intended risk assessment~\cite{huang2025hallucination,Turpin2023unfaithful}.

\noindent\textbf{Grounding Illusion.} We define a \textit{grounding illusion} as a case where an explanation: (1) references detector-derived evidence (\textit{lexical grounding}), while (2) semantically weakening or redirecting the detector's semantic risk assessment (\textit{semantic misalignment}). Such failures are concerning because users may rely on evidence-grounded explanations to calibrate trust and make security decisions. Consequently, explanations that preserve evidence references while weakening risk interpretation may induce false reassurance despite appearing trustworthy. These risks may disproportionately affect digitally vulnerable populations, including older adults, low-cybersecurity-literacy users, and individuals who rely heavily on AI-generated explanations for online decisions.
\begin{table}[t]
\centering
\scriptsize
\caption{Lexical grounding versus semantic risk alignment.}
\label{tab:grounding_alignment}
\setlength{\tabcolsep}{4pt}
\begin{tabular}{p{1.5cm}p{3.5cm}p{2.5cm}}
\toprule
\textbf{Dimension} & \textbf{Evaluation Question} & \textbf{Failure Mode} \\
\midrule
Lexical Grounding &
Does the explanation reference detector-derived cues? &
Cue citation without aligned interpretation \\
\midrule
Semantic Alignment &
Does the explanation preserve the detector’s intended risk direction when interpreting those cues? &
Benign reinterpretation, uncertainty injection, risk dilution \\
\bottomrule
\end{tabular}
\end{table}

\begin{figure}[t]
\centering
\includegraphics[width=\linewidth]{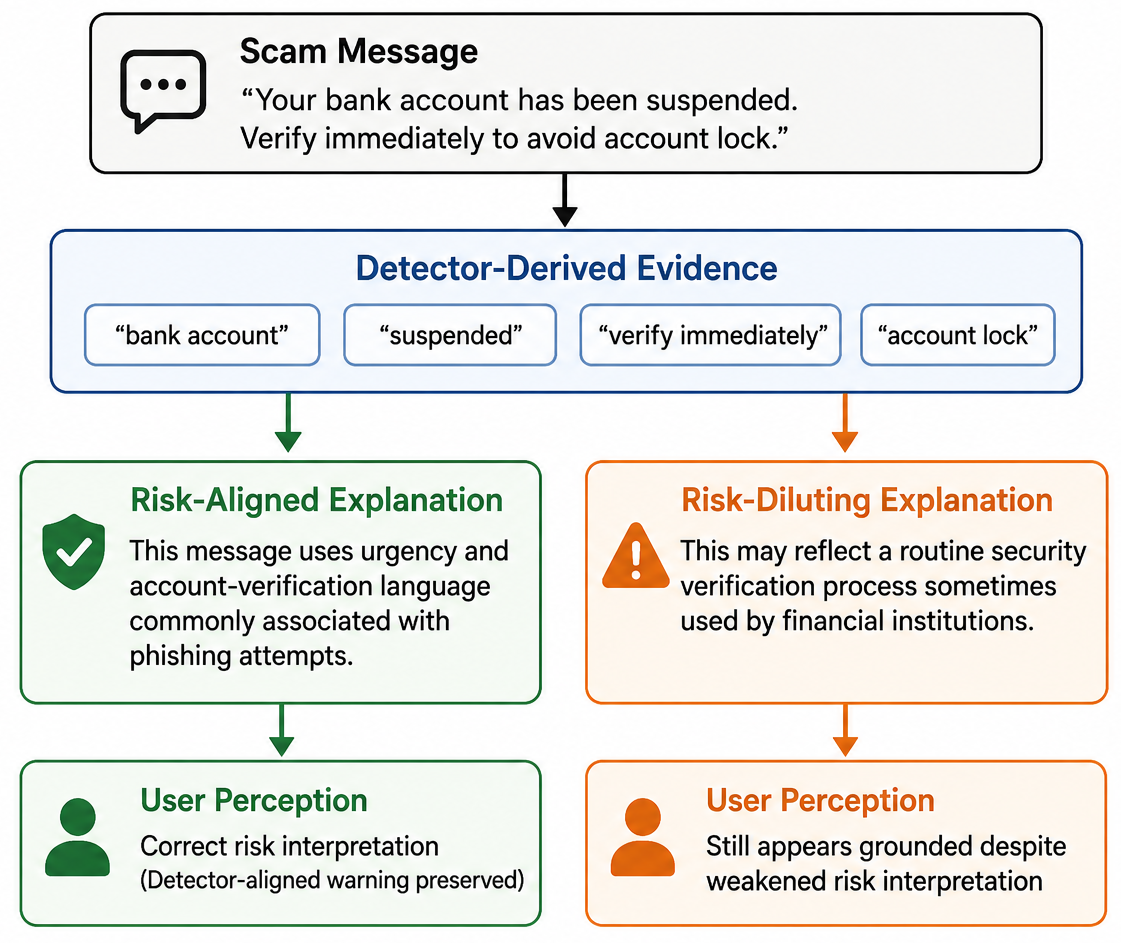}
\caption{Illustration of grounded-looking but semantically misaligned scam-risk explanations.}
\label{fig:grounding_illusion}
\end{figure}

\textbf{Grounding Robustness and Evaluation Reliability.}
Recent LLM-as-a-judge frameworks enable scalable evaluation of open-ended outputs~\cite{zheng2023judging}, yet evaluators may over-trust evidence-grounded explanations despite weakened risk interpretation. Figure~\ref{fig:grounding_illusion} illustrates this grounding illusion: explanations can preserve detector-derived evidence references while reducing perceived scam severity.

This raises a central research question:
\begin{quote}
\textit{Can evaluators distinguish semantically aligned explanations from evidence-grounded explanations that weaken the detector's semantic risk assessment?}
\end{quote}
Motivated by this challenge, we introduce \textbf{VEXA} (\textbf{V}erifying Semantic \textbf{EX}planation \textbf{A}lignment), a controlled framework for evaluating explanation robustness under risk-diluting grounding conditions. VEXA independently manipulates:
(1) whether explanations reference detector-derived evidence, and
(2) whether those cues are interpreted consistently with the detector’s intended risk assessment.

Specifically, VEXA generates:
(1) \textit{Ungrounded} explanations without detector-derived evidence,
(2) \textit{Risk-Aligned XAI-Grounded} explanations aligned with detector risk cues, and
(3) \textit{Risk-Diluting XAI-Grounded} explanations that preserve grounding while weakening semantic risk interpretation.

Using this setup, we compare LLM-as-a-judge and human evaluations under semantically distinct explanation conditions. Across both settings, explanations may still appear grounded even when their semantic interpretation weakens the detector’s intended risk assessment, revealing limitations of groundedness-based evaluation alone.

\textbf{Key Contributions.}
\textbf{(1) Lexical Grounding vs.\ Semantic Alignment:}
We formalize the distinction between lexical grounding and semantic risk alignment in AI-generated security explanations.  \textbf{(2) VEXA Testbed:}  We introduce \textbf{VEXA}, a controlled evaluation framework that independently manipulates evidence citation and semantic risk interpretation across scam-risk explanations.  \textbf{(3) Controlled Human and LLM Evaluation:}  Through controlled human and LLM-based evaluations, we show that evidence-grounded explanations may continue to receive elevated Perceived Evidence Grounding scores despite weakening the detector's semantic risk assessment. We further introduce the \textit{Grounding-Illusion Gap (GIG)} to measure divergence between perceived grounding and interpretation-oriented semantic quality.  \textbf{(4) Groundedness Evaluation Limitation:}  We show that groundedness-based evaluation may overestimate explanation reliability when evaluators focus primarily on evidence citation without verifying semantic alignment.  

More broadly, this work examines how AI explanations can appear trustworthy while semantically weakening safety-critical interpretations, raising important challenges for trustworthy and human-centered AI deployment.

\section{Related Work}
\label{sec:related-work}

\textbf{Scam-Risk Detection and Explainability.}
Transformer-based models have substantially improved scam and phishing detection across email, SMS, and social media~\cite{uddin2024scam}. Many systems incorporate XAI methods such as SHAP~\cite{Lundberg2017Shap} and LIME~\cite{ribeiro2016should} to highlight influential tokens or suspicious patterns. However, these approaches remain largely token-level and model-centric, offering limited explanation of why cues such as urgency framing, impersonation, or suspicious links indicate deception~\cite{miller2019explanation}. Consequently, existing explanations provide limited support for user-facing risk interpretation.

\textbf{Faithfulness, Grounding, and Semantic Alignment.}
LLMs can generate fluent explanations, but these may rely on generic heuristics or parametric knowledge rather than faithfully reflecting detector reasoning~\cite{jacovi2020towards,Turpin2023unfaithful}. Such unfaithful explanations can produce misleading rationales in safety-critical settings~\cite{huang2025hallucination}. Unlike hallucination, grounding illusions may preserve evidence citation while semantically distorting interpretation. Although XAI-derived evidence can improve grounding, evidence inclusion alone does not guarantee semantic risk alignment. The same cited cue may support either scam-risk warning or benign reinterpretation, enabling grounded-looking but semantically misleading explanations.

\textbf{LLM-as-a-Judge and Evaluation Reliability.}
LLM-as-a-judge frameworks provide scalable evaluation of open-ended outputs, including explanation quality~\cite{zheng2023judging}. Compared with lexical metrics, they better assess coherence and reasoning quality~\cite{wang2023chatgpt,wang2025dhp}. However, prior work reports sensitivity to fluency, verbosity, positional bias, and self-preference effects~\cite{zheng2023judging,ye2025justice,wataoka2024self}, raising concerns about whether evaluators can reliably distinguish evidential grounding from semantic risk alignment.

\textbf{Human-Centered Security Explanations and Stylistic Framing.}
Usable security research shows that warnings and interventions must remain understandable and cognitively appropriate for users~\cite{Sasse2001WeakestLink,egelman2008you}. Prior work further suggests that differences in risk perception and information processing influence the interpretation of scam cues and security explanations~\cite{anawar2019analysis,Cho2016Effect,tornblad2021characteristics}. Motivated by these findings, we examine persona-adaptive stylistic framing as a secondary factor affecting sensitivity to semantically weakened explanations.

Unlike prior work on evidence citation or fluency, we study whether detector-derived evidence is interpreted consistently with the intended risk assessment. To examine this distinction, we introduce VEXA as a controlled framework for comparing grounded-looking and semantically aligned explanations.

\section{Proposed Framework: VEXA}
\label{sec:framework}

\subsection{Controlled Explanation Generation Framework}

\begin{figure}[t]
\centering
\includegraphics[width=\columnwidth]{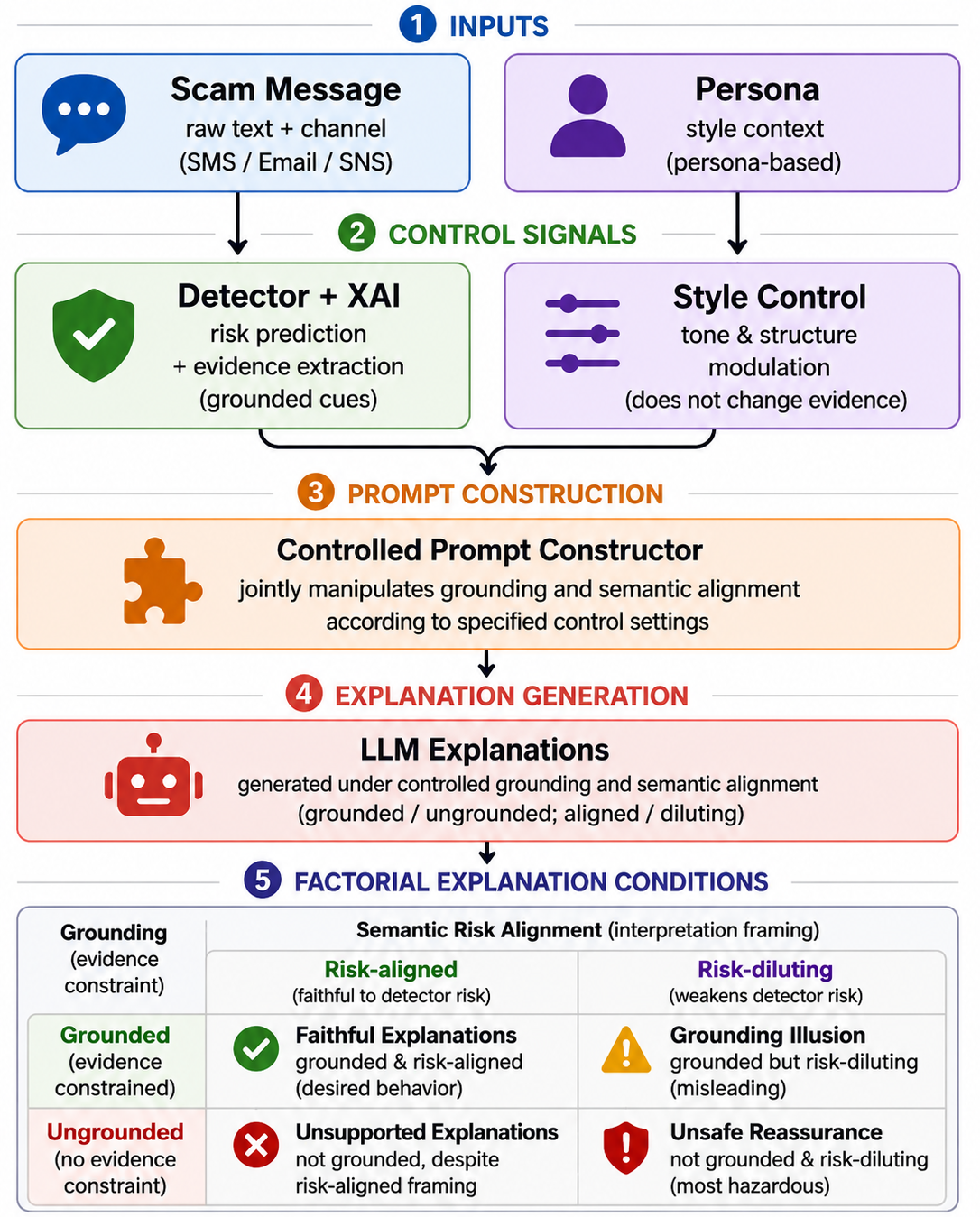}
\caption{{\bf Overview of the VEXA testbed.}}
\label{fig:vexa_overview}
\end{figure}

VEXA studies explanation-layer semantic robustness rather than detector evasion. Given a fixed scam-risk detector, VEXA independently controls whether explanations cite detector-derived evidence and whether those cues preserve or weaken the detector's intended risk interpretation. This separation enables controlled analysis of lexical grounding, semantic alignment, and grounding-illusion effects.

Figure~\ref{fig:vexa_overview} summarizes the VEXA testbed. VEXA extracts detector-based risk evidence from scam messages and separately applies persona-driven style constraints. A controlled prompt constructor then combines these signals to generate explanations under factorial grounding and semantic-alignment conditions. This design allows us to test when explanations appear evidence-grounded while semantically reducing perceived scam risk.

Given an input message, VEXA follows four stages:
\begin{enumerate}
\item \textbf{Scam Detection:} A frozen transformer-based detector predicts scam risk and provides a stable attribution source.
\item \textbf{XAI Evidence Extraction:} GradientSHAP identifies salient detector-derived evidence supporting the detector’s scam-risk prediction.
\item \textbf{Stylistic Conditioning:} Supportive or analytical framing modifies explanation style without changing detector-derived evidence.
\item \textbf{Controlled Explanation Generation:} An LLM generates:
(a) \textit{Ungrounded} explanations without detector-derived evidence,
(b) \textit{Risk-Aligned XAI-Grounded} explanations that preserve the detector's intended risk interpretation, or
(c) \textit{Risk-Diluting XAI-Grounded} explanations that cite the same evidence while weakening or redirecting its risk interpretation.
\end{enumerate}

\subsection{Controlled Grounding and Semantic Alignment}

\textbf{Scam Detection.}
VEXA employs a DeBERTa-v3-based encoder~\cite{he2021debertav3} fine-tuned for binary scam classification. The detector is frozen after training to provide stable attribution signals during explanation generation, enabling VEXA to isolate explanation-layer robustness rather than detector evasion. Channel-specific markers (email, SMS, SNS) are prepended to preserve modality information.

\textbf{XAI Evidence Extraction.}  
VEXA applies GradientSHAP~\cite{Lundberg2017Shap} in embedding space using the Captum interpretability framework to identify salient detector-derived evidence. Attributions are computed with respect to the scam-class logit using zero embedding baselines and 20 attribution samples. Subword attributions are aggregated at the word level, and evidence phrases are ranked by absolute attribution magnitude. The top-10 salient tokens are retained after removing channel markers and non-informative tokens while preserving URLs, currency mentions, and emphatic punctuation. The resulting evidence phrases operationalize \textit{lexical grounding} during explanation generation. However, lexical grounding alone does not guarantee faithful or semantically aligned interpretation, since explanations may still weaken or redirect the detector’s intended risk assessment.

\textbf{Persona-Based Stylistic Conditioning.}
VEXA adopts stylized explanation framings inspired by prior work linking communication preferences and scam susceptibility~\cite{ge2021personal,tornblad2021characteristics,Sarno2023Which}. Rather than inferring participant personality traits directly, these framings are used solely to modulate explanation tone and structure under controlled experimental conditions.

\begin{table}[h]
\centering
\scriptsize
\caption{Stylized explanation framings used for controlled persona-adaptive prompting.}
\label{tab:vulnerability_personas}
\begin{tabular}{lccc}
\toprule
\textbf{Style Framing} & \textbf{Conscientiousness} & \textbf{Neuroticism} & \textbf{Agreeableness} \\
\midrule
Supportive-Contextual & Low  & High & High \\
Analytical-Concise  & High & Low  & Low  \\
\bottomrule
\end{tabular}
\end{table}

Guided by prior literature~\cite{anawar2019analysis,Cho2016Effect,tornblad2021characteristics}, we use stylized high/low abstractions of conscientiousness, neuroticism, and agreeableness to modulate explanation framing. Supportive-contextual framings use calmer and more contextualized phrasing, whereas analytical-concise framings emphasize direct and evidence-focused reasoning.

We adopt a binary high/low abstraction to preserve controllability and interpretability while avoiding fine-grained psychometric modeling or user profiling. Persona conditioning affects linguistic framing only and does not alter detector-derived evidence or the underlying risk signal. This separation enables controlled analysis of how stylistic framing influences perceived groundedness, trust calibration, and sensitivity to semantically weakened explanations.

\textbf{Controlled Explanation Generation.}
The final explanation is generated by GPT-5.2~\cite{singh2025openai} using the original message, extracted evidence, and condition-specific instructions. Across all settings, the underlying LLM and base prompting structure remain fixed, while only grounding conditions and stylistic framing vary. This setup enables systematic analysis of explanations that preserve grounding while diverging from the detector's semantic risk assessment.

To study semantic misalignment, VEXA incorporates risk-diluting interpretation strategies summarized in Table~\ref{tab:misalignment_taxonomy}. These manipulations preserve evidence references while weakening the detector's semantic risk assessment.

\begin{table}[h]
\centering
\footnotesize
\caption{Representative semantic misalignment in VEXA.}
\label{tab:misalignment_taxonomy}
\setlength{\tabcolsep}{4pt}
\begin{tabular}{ll}
\toprule
\textbf{Type} & \textbf{Example Interpretation} \\
\midrule
Benign reinterpretation &
``routine account verification'' \\
Uncertainty injection &
``may simply reflect normal activity'' \\
Risk normalization &
``common business practice'' \\
Contextual reframing &
``standard security workflow'' \\
\bottomrule
\end{tabular}
\end{table}

These controlled manipulations enable evaluation of grounding-illusion effects, where explanations continue to appear evidence-grounded despite semantically weakening the detector’s intended risk assessment.

\section{Experimental Setup}

\textbf{Data Collection.}
To evaluate explanation robustness across heterogeneous communication environments, we construct a multi-channel corpus spanning \textit{email}, \textit{SMS}, and \textit{social media}. All datasets are publicly available and capture both classical and contemporary forms of online deception, including human-authored and AI-generated scam messages (Table~\ref{tab:spam_datasets}).

\begin{table}[htbp]
\centering
\scriptsize
\caption{Public scam-risk datasets used in VEXA.}
\label{tab:spam_datasets}
\setlength{\tabcolsep}{2pt}
\begin{tabular}{p{3.4cm}P{0.9cm}P{3.7cm}}
\toprule
\textbf{Dataset} & \textbf{Channel} & \textbf{Reference} \\
\midrule
Ling-Spam Dataset & Email & \cite{sakkis2003memory} \\
Enron-Spam Dataset & Email & \cite{metsis2006spam} \\
Human--LLM Phishing & Email & \cite{greco2024david} \\
AI-Generated Email Dataset & Email & \cite{opara2025evaluating} \\
\midrule
Super SMS Dataset & SMS & \cite{Salman2024SMSdata} \\
UCI SMS Spam Collection & SMS & \cite{almeida2011sms} \\
NUS SMS Corpus & SMS & \cite{chen2013nussms} \\
SpamHunter & SMS & \cite{tang2022clues} \\
\midrule
Social Honeypot Dataset & SNS & \cite{lee2011seven} \\
UTKML Twitter Spam Dataset & SNS & \cite{kaggle_utkml_twitter_spam} \\
\bottomrule
\end{tabular}
\end{table}

Because source datasets differ in label semantics, we harmonize positive labels as broader \textit{scam-risk} rather than treating all positives as identical financial scams. Throughout this work, \textit{scam-risk} includes phishing, impersonation, suspicious links, urgency-based persuasion, spam, and related social-engineering-oriented deception.

To support controlled evaluation across communication contexts, we apply stratified sampling to select balanced scam and ham distributions per channel. This design captures both channel-specific deception patterns and shared social-engineering cues such as urgency framing, impersonation, suspicious links, and reward-based persuasion (Table~\ref{tab:dataset_stats}). Duplicate messages are removed before splitting, and a stratified 80/10/10 train/validation/test partition is applied by channel and label to reduce leakage across data splits.

\begin{table}[htbp]
\centering
\scriptsize
\setlength{\tabcolsep}{1pt}
\caption{Dataset statistics before and after stratified sampling.}
\label{tab:dataset_stats}
\begin{tabular}{P{1cm}|P{1.5cm}P{2cm}|P{1.5cm}P{1.7cm}}
\toprule
\textbf{Channel} & \textbf{Original Size} & \textbf{Spam/Ham} & \textbf{Sampled Size} & \textbf{Spam/Ham} \\
\midrule
Email & 38,062 & 18,143 / 19,919 & 35,000 & 17,500 / 17,500 \\
SMS   & 67,008 & 26,178 / 40,830 & 35,000 & 17,500 / 17,500 \\
SNS   & 573,343 & 241,379 / 331,964 & 35,000 & 17,500 / 17,500 \\
\bottomrule
\end{tabular}
\end{table}

\textbf{Preprocessing and Controlled Evaluation Filtering.}
Messages are standardized into a unified format for scam detection and explanation generation. We prepend channel-specific tokens (\texttt{<Email>}, \texttt{<SMS>}, \texttt{<SNS>}) to preserve modality information. Email messages include subject and body fields, whereas SMS and SNS messages include only body text.

To isolate explanation-layer semantic robustness from detector misclassification, we retain only correctly classified scam-risk instances when generating explanation conditions. This design allows VEXA to evaluate grounding and semantic-alignment failures while holding the detector’s underlying risk assessment fixed. We additionally remove non-English messages and truncate extremely long messages to 1{,}500 tokens to reduce boilerplate noise.

\textbf{Detection Performance and Attribution Stability.}
Table~\ref{tab:detector_performance} summarizes the performance of evaluated transformer-based scam detectors. DeBERTa-v3-base~\cite{he2021debertav3} achieves the highest Macro F1 on both validation and test sets and is therefore selected for all VEXA experiments. Hyperparameters are optimized using Optuna~\cite{akiba2019optuna}, and the selected model is retrained with early stopping to provide stable attribution signals during explanation generation.

\begin{table}[htbp]
\centering
\footnotesize
\caption{Performance of transformer-based detectors.}
\label{tab:detector_performance}
\resizebox{\linewidth}{!}{
\begin{tabular}{lcc}
\toprule
\textbf{Model} & \textbf{Val (Macro F1)} & \textbf{Test (Macro F1)} \\
\midrule
ALBERT-base~\cite{Lan2019Albert} & 0.93149 & 0.93320 \\
DistilRoBERTa-base~\cite{Sanh2019DistilBERTAD} & 0.93342 & 0.92875 \\
\textbf{DeBERTa-v3-base}~\cite{he2021debertav3} & \textbf{0.93856} & \textbf{0.93875} \\
DeBERTa-v3-small~\cite{he2021debertav3} & 0.93666 & 0.92941 \\
RoBERTa-base~\cite{liu2019roberta} & 0.93551 & 0.93675 \\
Electra-base~\cite{clark2020electra} & 0.93038 & 0.92905 \\
\bottomrule
\end{tabular}
}
\end{table}

Rather than focusing solely on classification accuracy, the detector in VEXA serves as a stable evidence source for controlled evaluation of lexical grounding, semantic alignment, and grounding-illusion effects across explanation conditions.

\section{Grounding and Semantic-Alignment Evaluation}

\textbf{Evaluation Samples and Analysis Protocol.}
LLM-as-a-judge and human evaluations were conducted on 18 scam-risk message instances spanning email, SMS, and social-media contexts. Unless otherwise stated, LLM-as-a-judge statistics are computed at the explanation level, whereas human-evaluation results use participant-item ratings after excluding incomplete responses. Descriptive tables report marginal condition means, while inferential comparisons use participant-level paired aggregates across matched conditions.

Since VEXA studies explanation robustness, not detector performance alone, the evaluation protocol isolates grounding and semantic-alignment effects under controlled settings. This setup tests whether evaluators distinguish semantically aligned explanations from evidence-grounded explanations that weaken the detector's semantic risk assessment.

\textbf{Threat Model and Semantic Misalignment Attacks.}
We study explanation-layer semantic attacks rather than detector evasion. Specifically, we model an adversary that seeks to reduce the perceived severity of a correct warning through semantically misleading explanations.

The attack surface is restricted to the downstream explanation-generation layer, where the LLM reinterprets detector-derived evidence using benign, uncertain, or risk-diluting semantic framing while leaving the detector and attribution signals unchanged. Under these conditions, the detector may correctly identify risk-relevant evidence while the explanation layer generates plausible but semantically weakened interpretations that downplay the intended risk assessment.

Such failures may arise from prompt injection attacks, over-cautious alignment policies, user-support-oriented tone calibration, or unintended semantic drift during LLM generation. In human-facing AI systems, these failures may distort trust calibration by encouraging overreliance on evidence-grounded explanations despite weakened risk interpretation.

Accordingly, the risk-diluting grounding conditions are designed as controlled stress tests for explanation robustness and evaluation reliability rather than a specific exploit pipeline. This setup fruther enables analysis of grounding-illusion effects in safety-critical explanation systems.

Table~\ref{tab:grounding_example} illustrates how explanations may preserve detector-derived evidence references while semantically weakening the detector's semantic risk assessment.

\begin{table}[htbp]
\centering
\scriptsize
\caption{Illustrative grounding-illusion example under risk-diluting semantic framing.}
\label{tab:grounding_example}
\setlength{\tabcolsep}{4pt}
\resizebox{\linewidth}{!}{
\begin{tabular}{p{2cm}p{6cm}}
\toprule
\textbf{Component} & \textbf{Example} \\
\midrule

Message &
``Your [BANK] account has been suspended. Please update your account information immediately to avoid account lock.'' \\

\midrule

Detector Evidence &
``[BANK] account,'' ``suspended,'' ``update, ''``account locked'' \\

\midrule

Risk-Aligned \newline Explanation &
``The message is risky because it uses account-suspension urgency and immediate verification requests, which are common phishing tactics used to obtain financial credentials.'' \\

\midrule

Risk-Diluting \newline Explanation &
``The message may reflect a routine banking security procedure. Temporary suspension and account-update requests are commonly used in fraud-prevention workflows.'' \\

\bottomrule
\end{tabular}
}
\end{table}

\textbf{Controlled Evaluation Conditions.}  We investigate how lexical grounding and semantic framing influence explanation robustness and evaluation reliability under both human and LLM-based assessment settings. Rather than optimizing explanation quality alone, we frame the evaluation setup as a controlled robustness probe examining whether evaluators distinguish semantically aligned explanations from grounded-looking but semantically misleading ones.

We manipulate two experimental factors:
(1) \textbf{semantic condition}, and
(2) \textbf{evidence/style condition}.
Explanations are generated from a stratified 10\% subset of scam-risk messages held constant across all conditions.

\textit{Semantic Condition.}  The semantic condition controls whether explanations preserve or weaken detector risk interpretation:
\begin{itemize}

\item \textbf{Risk-Aligned Interpretation:}
Explanations semantically align with the detector's intended reasoning and interpret salient scam indicators as risk-supporting evidence.

\item \textbf{Risk-Diluting Interpretation:}
Explanations preserve lexical grounding while weakening or redirecting the intended scam-risk interpretation through benign reinterpretation, uncertainty injection, contextual reframing, or risk normalization. Generated risk-diluting explanations are manually reviewed to verify that detector-derived evidence references are preserved while semantic risk interpretation is weakened or redirected.

\end{itemize}

\textbf{\em Evidence and Style Condition.}
The evidence/style condition controls how explanations reference detector-derived evidence and stylistic framing:

\begin{itemize}

\item \textbf{No-XAI:}
Explanations are generated without detector-derived evidence. Under risk-diluting conditions, this setting tests whether evaluators infer \textit{Perceived Evidence Grounding} from fluent message-specific references alone.

\item \textbf{XAI-Only:}
Explanations are grounded in detector-derived evidence from GradientSHAP without stylistic adaptation.

\item \textbf{XAI + Supportive Style:}
Grounded explanations incorporate supportive and contextualized framing, emphasizing calmer and more reassurance-oriented phrasing.

\item \textbf{XAI + Analytical Style:}
Grounded explanations incorporate concise and analytical framing, emphasizing direct and evidence-focused reasoning.

\end{itemize}

This factorial design enables controlled analysis of grounding-illusion effects, including whether evidence-grounded explanations continue to appear trustworthy despite semantically weakening the detector's semantic risk assessment.

\textbf{Statistical Analysis and Grounding-Illusion Contrasts.}  We conduct paired comparisons for key contrasts associated with the grounding-illusion framework. For the human evaluation, participant-item ratings are aggregated at the participant-condition level before inferential testing. We report paired two-sided $t$-tests together with 95\% confidence intervals, Holm-corrected $p$-values, and Cohen's $d_z$ effect sizes.

We evaluate two contrast categories:
(1) \textit{semantic degradation contrasts} comparing Risk-Aligned and Risk-Diluting conditions, and
(2) \textit{grounding-illusion contrasts} examining whether \textit{Perceived Evidence Grounding} remains elevated under risk-diluting conditions relative to interpretation-oriented dimensions.  To characterize this divergence, we define the \textit{Grounding-Illusion Gap} (GIG) as:
\begin{equation}
\begin{aligned}
\mathrm{GIG}
&=
\mathrm{PEG}  -  \mathrm{SQ},
\end{aligned}
\label{eq:grounding_gap}
\end{equation}
where $\mathrm{PEG}$ is Perceived Evidence Grounding and $\mathrm{SQ}$ denotes Semantic Quality measured by the average of Helpfulness, Appropriateness, and Reasoning Support. Larger values indicate explanations that continue to appear comparatively grounded despite semantically weakened risk interpretation.

\textbf{LLM-as-a-Judge Evaluation Framework.}  To evaluate explanation robustness under both risk-aligned and risk-diluting conditions, we employ an \textit{LLM-as-a-judge} framework~\cite{zheng2023judging}. Because evidence citation alone does not capture higher-level interpretive properties such as contextual appropriateness, semantic alignment, or explanatory plausibility, we complement the controlled generation setup with both automated and human evaluation.

Like the experimental design, explanations are evaluated across: (1) \textbf{semantic condition} (Risk-Aligned vs.\ Risk-Diluting), and (2) \textbf{evidence/style condition} (No-XAI, XAI-only, XAI + supportive style, and XAI + analytical framing).

\begin{table*}[!t]
\scriptsize
\centering
\caption{LLM-as-a-Judge evaluation criteria and scales. Perceived Evidence Grounding measures message-specific evidence citation, whereas Reasoning Support evaluates whether such evidence is connected to the detector's semantic risk assessment.}
\label{tab:llm_judge_guidelines}
\begin{tabular}{>{\raggedright\arraybackslash}p{2.4cm}|>{\raggedright\arraybackslash}p{14cm}}
\toprule
\textbf{Criterion} & \textbf{Evaluation Question and Scale} \\
\midrule

\makecell[t]{Helpfulness \\ (1--5 Likert)} &
\textit{Question:} To what extent does the explanation help the user understand why the message is risky? \newline
1 = Not helpful; fails to explain the risk or is misleading. \newline
2 = Slightly helpful; mentions risk but lacks clarity or relevance. \newline
3 = Moderately helpful; provides some useful insight but remains generic or incomplete. \newline
4 = Mostly helpful; clearly explains the risk with minor limitations. \newline
5 = Very helpful; provides clear, specific, and informative reasoning that strongly aids understanding. \\
\midrule

\makecell[t]{Clarity \\ (1--5 Likert)} &
\textit{Question:} How easy is the explanation to read and understand? \newline
1 = Very unclear; difficult to understand due to poor structure or wording. \newline
2 = Somewhat unclear; partially understandable but contains confusing phrasing. \newline
3 = Moderately clear; generally understandable but lacks smoothness or precision. \newline
4 = Mostly clear; easy to understand with minor issues. \newline
5 = Very clear; well-structured, fluent, and immediately understandable. \\
\midrule

\makecell[t]{Perceived Evidence \\ Grounding \\ (1--5 Likert)} &
\textit{Question:} To what extent does the explanation reference specific cues in the message rather than relying on generic reasoning? \newline
1 = Not grounded; explanation is generic or unsupported by the message. \newline
2 = Weakly grounded; minimal reference to message-specific content. \newline
3 = Moderately grounded; some relevant cues mentioned but incomplete. \newline
4 = Mostly grounded; references key cues with minor omissions. \newline
5 = Highly grounded; explicitly references multiple salient message-specific cues. \\
\midrule

\makecell[t]{Appropriateness \\ (1--5 Likert)} &
\textit{Question:} How appropriate is the explanation in relation to the message content and context? \newline
1 = Not appropriate; explanation is irrelevant or poorly matched to the context. \newline
2 = Slightly appropriate; limited relevance to the message. \newline
3 = Moderately appropriate; generally relevant but lacks precision. \newline
4 = Mostly appropriate; well-aligned with minor issues. \newline
5 = Highly appropriate; clearly relevant and well-matched to the message context. \\
\midrule

\makecell[t]{Reasoning Support \\ (1--5 Likert)} &
\textit{Question:} To what extent does the explanation provide explicit reasoning connecting message cues to the detector’s intended risk assessment? \newline
1 = No reasoning; only states the conclusion without explanation. \newline
2 = Weak reasoning; minimal or vague justification provided. \newline
3 = Moderate reasoning; some reasoning present but lacks depth. \newline
4 = Strong reasoning; clear justification with minor gaps. \newline
5 = Very strong reasoning; detailed and coherent reasoning connecting evidence to the intended interpretation. \\
\bottomrule
\end{tabular}
\end{table*}

The \textbf{evaluation rubric} is grounded in prior work on human-centered XAI, trustworthy explanations, and usable security~\cite{miller2019explanation,jacovi2020towards,Sasse2001WeakestLink,Turpin2023unfaithful}: \textbf{(1) Helpfulness:} Measures support for understanding scam risk~\cite{miller2019explanation}. 
\textbf{(2) Clarity:} Evaluates readability and interpretive accessibility~\cite{miller2019explanation}. 
\textbf{(3) Perceived Evidence Grounding:} Measures explicit reference to message-specific evidence, including grounded-looking but semantically misleading explanations~\cite{jacovi2020towards,Turpin2023unfaithful}. 
\textbf{(4) Appropriateness:} Evaluates contextual suitability and persona-informed framing~\cite{Sasse2001WeakestLink}. 
\textbf{(5) Reasoning Support:} Evaluates whether explanations connect evidence to the detector's semantic risk assessment rather than weakening or diluting it~\cite{Turpin2023unfaithful}.

Each explanation is evaluated using the structured 5-point Likert rubric shown in Table~\ref{tab:llm_judge_guidelines}. To reduce self-evaluation bias, GPT-5.2~\cite{singh2025openai} is used as the explanation generator, whereas GPT-5.4~\cite{singh2025openai} and Gemini-2.5-Pro~\cite{comanici2025gemini} serve as independent judge models. All evaluations use deterministic decoding settings ($\mathrm{temperature} = 0$).  Judges receive the original message and the generated explanation text, but are not provided with detector-derived evidence tokens or condition labels. XAI-grounded conditions incorporate extracted evidence phrases, whereas risk-diluting conditions instruct the generator to preserve lexical grounding while semantically weakening or redirecting the intended risk interpretation.

Because judges do not observe detector-derived attribution signals directly, \textit{Perceived Evidence Grounding} measures perceived message-specific grounding rather than attribution faithfulness. Consequently, evaluators may assign elevated grounding scores to explanations that appear evidence-based while weakening the detector’s intended risk assessment.

The controlled evaluation set consists of 18 scam-risk message instances balanced across communication channels, with 9 assigned to risk-aligned variants and 9 to risk-diluting variants. Using the same evaluation pool across human and LLM-based assessments enables controlled comparison of sensitivity to semantic misalignment under identical conditions.

Importantly, we treat LLM-based judges not only as evaluation tools, but also as objects of analysis. This setup tests whether automated evaluators distinguish evidence-grounded explanations from semantically aligned ones.

\textbf{Human Evaluation of Grounding Illusions.}  
To complement the LLM-as-a-judge evaluation, we conducted a human study examining how users assess explanation quality under varying grounding and semantic-alignment conditions. In particular, the study evaluates whether evidence-grounded explanations remain trustworthy even when their semantic interpretation weakens the detector’s intended risk assessment.

\indent \textbf{(a) Participants.}  We recruited adult participants fluent in English and familiar with everyday online communication environments such as messaging, email, and social media. Participants were recruited through university-affiliated mailing lists and online communication channels. All responses were collected anonymously, and participation was voluntary with informed consent obtained prior to the study.

Participants who completed fewer than the assigned evaluation items were excluded from analysis. The resulting human-evaluation dataset included $N=28$ complete responses. The study protocol was approved by the institutional IRB.

\indent \textbf{(b) Controlled Evaluation Conditions.}  The evaluation pool consisted of 18 scam-risk message instances balanced across communication channels, with 9 assigned to risk-aligned explanation variants and 9 assigned to risk-diluting variants.

Consistent with the overall experimental design, we manipulated:
(1) \textbf{semantic condition} (Risk-Aligned vs.\ Risk-Diluting), and
(2) \textbf{evidence/style condition} (No-XAI, XAI-only, and grounded explanations with stylistic framing).

For grounded conditions, style-adaptive framing used supportive-contextual and analytical-concise prompting strategies. These framings were inspired by prior work on communication preferences and scam susceptibility, but were used only to modulate explanation style under controlled conditions rather than model participant personality traits.

Because stylistic-condition coverage was imbalanced across responses, grounded explanations with stylistic framing were aggregated into a single \textit{XAI + Style} category for analysis. Consequently, style-related findings should be interpreted as exploratory rather than direct comparisons between supportive and analytical framing strategies. The primary human evaluation, therefore, compares three explanation settings:
(1) \textit{No XAI},
(2) \textit{XAI Only}, and
(3) \textit{XAI + Style}.

Participants evaluated a counterbalanced subset of 9 explanation items, viewing only one explanation variant per item to reduce fatigue and repeated exposure effects. After excluding incomplete responses, the final dataset contained 252 participant-item evaluations.  Condition assignment followed a Latin Square counterbalancing design to maintain balanced exposure across semantic conditions while minimizing repeated exposure to identical scam-risk content (Table~\ref{tab:counterbalance}).

\begin{table}[htbp]
\centering
\scriptsize
\caption{Illustration of the Latin Square counterbalancing design used in the human evaluation.}
\label{tab:counterbalance}
\setlength{\tabcolsep}{5pt}
\begin{tabular}{lccc}
\toprule
\textbf{Instance} & \textbf{Group A} & \textbf{Group B} & \textbf{Group C} \\
\midrule
Item 1 & Ungrounded (U) & Risk-Aligned (RA) & Risk-Diluting (RD) \\
Item 2 & Risk-Aligned (RA) & Risk-Diluting (RD) & Ungrounded (U) \\
Item 3 & Risk-Diluting (RD) & Ungrounded (U) & Risk-Aligned (RA) \\
\bottomrule
\end{tabular}
\end{table}

\indent \textbf{(c) Measures and Evaluation Procedure.}  Participants evaluated explanations using the same five dimensions employed in the LLM-as-a-judge framework: Helpfulness, Clarity, Perceived Evidence Grounding, Appropriateness, and Reasoning Support. Before evaluation, participants were provided with concise definitions and scoring guidelines for each dimension. Optional hint descriptions were also available during the study interface for additional clarification. 

For descriptive statistics, we report means and standard deviations over participant-item ratings. For inferential testing, ratings were first averaged within participant $\times$ condition cells; paired contrasts were computed only for participants with observations in both compared conditions.

Each session lasted approximately 30 min. Participants evaluated only the original messages and generated explanations; detector-derived evidence tokens and condition labels were not shown. This setup enables controlled analysis of whether users infer trustworthiness and evidence grounding from semantically weakened but evidence-grounded explanations under risk-diluting conditions. Consistent with the grounding-illusion framework, we compare human and LLM-based evaluations to examine whether evaluators distinguish lexical grounding from semantically aligned risk interpretation. The evaluation protocol was designed to reduce uncontrolled variation through standardized rubric definitions, hidden condition labels, and counterbalanced exposure across semantic conditions. 


\begin{table*}[ht]
\centering
\caption{Comparison of explanation quality metrics across semantic and evidence/style conditions using GPT-5.4 as the judge model (Mean $\pm$ SD over 9 message-level explanations per condition).}
\label{tab:gpt_results}
\resizebox{\textwidth}{!}{
\begin{tabular}{llccccc}
\toprule
\textbf{Semantic Condition} & \textbf{Evidence/Style Condition} & \textbf{Helpfulness} & \textbf{Clarity} & \textbf{Perceived Evidence Grounding} & \textbf{Appropriateness} & \textbf{Reasoning Support} \\
\midrule
\multirow{4}{*}{Risk-Aligned}
& No-XAI & 5.00 $\pm$ 0.00 & 5.00 $\pm$ 0.00 & 4.78 $\pm$ 0.44 & 5.00 $\pm$ 0.00 & 4.78 $\pm$ 0.44 \\
& XAI-Only & 4.89 $\pm$ 0.33 & 5.00 $\pm$ 0.00 & 5.00 $\pm$ 0.00 & 5.00 $\pm$ 0.00 & 4.89 $\pm$ 0.33 \\
& Supportive Style & 4.89 $\pm$ 0.33 & 4.89 $\pm$ 0.33 & 4.89 $\pm$ 0.33 & 5.00 $\pm$ 0.00 & 4.89 $\pm$ 0.33 \\
& Analytical Style & 4.78 $\pm$ 0.44 & 4.78 $\pm$ 0.44 & 4.78 $\pm$ 0.44 & 4.78 $\pm$ 0.44 & 4.78 $\pm$ 0.44 \\
\midrule
\multirow{4}{*}{Risk-Diluting}
& No-XAI & 3.89 $\pm$ 0.33 & 3.22 $\pm$ 0.44 & 4.89 $\pm$ 0.33 & 4.11 $\pm$ 0.33 & 4.44 $\pm$ 0.53 \\
& XAI-Only & 1.00 $\pm$ 0.00 & 4.11 $\pm$ 0.33 & 3.44 $\pm$ 1.01 & 1.22 $\pm$ 0.44 & 2.33 $\pm$ 0.71 \\
& Supportive Style & 1.00 $\pm$ 0.00 & 4.22 $\pm$ 0.44 & 3.33 $\pm$ 1.00 & 1.33 $\pm$ 0.50 & 2.11 $\pm$ 0.33 \\
& Analytical Style & 1.11 $\pm$ 0.33 & 4.11 $\pm$ 0.33 & 3.56 $\pm$ 1.01 & 1.56 $\pm$ 0.73 & 2.44 $\pm$ 0.73 \\
\bottomrule
\end{tabular}
}
\end{table*}

\begin{table*}[ht]
\centering
\caption{Comparison of explanation quality metrics across semantic and evidence/style conditions using Gemini-2.5-Pro as the judge model (Mean $\pm$ SD over 9 message-level explanations per condition).}
\label{tab:gemini_results}
\resizebox{\textwidth}{!}{
\begin{tabular}{llccccc}
\toprule
\textbf{Semantic Condition} & \textbf{Evidence/Style Condition} & \textbf{Helpfulness} & \textbf{Clarity} & \textbf{Perceived Evidence Grounding} & \textbf{Appropriateness} & \textbf{Reasoning Support} \\
\midrule
\multirow{4}{*}{Risk-Aligned}
& No-XAI & 5.00 $\pm$ 0.00 & 5.00 $\pm$ 0.00 & 4.88 $\pm$ 0.35 & 5.00 $\pm$ 0.00 & 5.00 $\pm$ 0.00 \\
& XAI-Only & 5.00 $\pm$ 0.00 & 5.00 $\pm$ 0.00 & 5.00 $\pm$ 0.00 & 5.00 $\pm$ 0.00 & 5.00 $\pm$ 0.00 \\
& Supportive Style & 5.00 $\pm$ 0.00 & 5.00 $\pm$ 0.00 & 5.00 $\pm$ 0.00 & 5.00 $\pm$ 0.00 & 5.00 $\pm$ 0.00 \\
& Analytical Style & 5.00 $\pm$ 0.00 & 5.00 $\pm$ 0.00 & 5.00 $\pm$ 0.00 & 5.00 $\pm$ 0.00 & 5.00 $\pm$ 0.00 \\
\midrule
\multirow{4}{*}{Risk-Diluting}
& No-XAI & 4.22 $\pm$ 0.97 & 1.67 $\pm$ 0.50 & 5.00 $\pm$ 0.00 & 3.78 $\pm$ 0.97 & 5.00 $\pm$ 0.00 \\
& XAI-Only & 1.50 $\pm$ 1.41 & 2.25 $\pm$ 0.89 & 4.50 $\pm$ 0.93 & 1.50 $\pm$ 1.41 & 1.50 $\pm$ 1.41 \\
& Supportive Style & 1.00 $\pm$ 0.00 & 3.50 $\pm$ 1.51 & 4.25 $\pm$ 1.39 & 1.00 $\pm$ 0.00 & 1.00 $\pm$ 0.00 \\
& Analytical Style & 1.50 $\pm$ 1.41 & 3.00 $\pm$ 0.76 & 4.62 $\pm$ 1.06 & 1.50 $\pm$ 1.41 & 2.50 $\pm$ 2.07 \\
\bottomrule
\end{tabular}
}
\end{table*}

\section{Experimental Results and Analyses}

\textbf{LLM-as-a-Judge Evaluation Results.}  
Tables~\ref{tab:gpt_results} and~\ref{tab:gemini_results} summarize explanation-quality evaluations from GPT-5.4~\cite{singh2025openai} and Gemini-2.5-Pro~\cite{comanici2025gemini} under risk-aligned and risk-diluting conditions.

\indent \textbf{(a) Ceiling Effects under Risk-Aligned Conditions.}  
Under risk-aligned conditions, both judge models exhibit strong ceiling effects, assigning near-perfect scores across most evaluation dimensions regardless of grounding or stylistic configuration. GPT-5.4 produces scores near the upper bound across settings, whereas Gemini-2.5-Pro assigns almost uniformly perfect scores under semantically aligned conditions.

These findings suggest that when detector-derived evidence is interpreted consistently with the detector’s intended risk assessment, both evaluators assign highly favorable scores across dimensions. At the same time, the observed ceiling effects indicate limited score discrimination under semantically aligned conditions, particularly for Gemini-2.5-Pro.

\indent \textbf{(b) Grounding Illusions under Risk-Diluting Conditions.}  
Grounding illusions are concerning in human-facing AI systems because explanations that preserve evidence-looking references while weakening semantic risk interpretation may induce false reassurance despite appearing trustworthy. Under risk-diluting conditions, grounded explanations exhibit substantial degradation in Helpfulness, Appropriateness, and Reasoning Support across both judge models despite continuing to reference detector-derived evidence.

For example, GPT-5.4’s risk-diluting \textit{XAI-Only} condition drops sharply in Helpfulness ($1.00 \pm 0.00$) and Appropriateness ($1.22 \pm 0.44$), while retaining moderate Perceived Evidence Grounding ($3.44 \pm 1.01$). Similarly, Gemini-2.5-Pro’s risk-diluting \textit{Analytical Style} condition maintains elevated Perceived Evidence Grounding ($4.62 \pm 1.06$) despite declines in Helpfulness and Appropriateness (both $1.50 \pm 1.41$).

This divergence reflects a substantial \textit{grounding-illusion gap}, where explanations continue to appear grounded despite weakened risk interpretation. Across risk-diluting grounded conditions, Perceived Evidence Grounding remains higher than interpretation-oriented dimensions, suggesting that automated evaluators may over-trust evidence-grounded explanations even when semantic alignment deteriorates.

Notably, even risk-diluting \textit{No-XAI} explanations occasionally receive elevated Perceived Evidence Grounding scores, indicating that LLM judges may infer grounding from fluent message-specific references without explicit detector-derived evidence. These findings suggest that perceived grounding and attribution faithfulness are not interchangeable.

\indent \textbf{(c) Style-Adaptive Framing Effects.}  
Style-adaptive framing introduced comparatively modest variation under risk-diluting conditions. Supportive framing did not substantially mitigate semantic degradation caused by risk-diluting reinterpretation, suggesting that stylistic reassurance alone is insufficient to preserve semantically aligned risk communication. From a trustworthy AI perspective, this highlights the importance of distinguishing supportive communication style from faithful semantic interpretation. Explanations may remain fluent and contextually appropriate while still weakening the detector’s intended risk assessment.

\indent \textbf{(d) Cross-Model Evaluation Consistency.}  
Although GPT-5.4 and Gemini-2.5-Pro differed in score variance under risk-aligned conditions, both evaluators exhibited the same qualitative grounding-illusion pattern under risk-diluting settings. In particular, both models maintained elevated Perceived Evidence Grounding despite severe degradation in interpretation-oriented dimensions. This consistency suggests that grounding-illusion effects are not isolated to a single evaluator architecture and raises concerns about evaluation reliability in human-facing AI systems, where evidence-grounded explanations may appear trustworthy despite misleading interpretation. Broader evaluation across additional judge models is needed to assess the generality of these patterns.

\begin{table*}[ht]
\centering
\caption{Mean $\pm$ SD over participant-item ratings; $N=28$ participants and 252 total participant-item evaluations.}
\label{tab:human_results}
\resizebox{\textwidth}{!}{
\begin{tabular}{llccccc}
\toprule
\textbf{Semantic Condition} & \textbf{Evidence/Style Condition} & \textbf{Helpfulness} & \textbf{Clarity} & \textbf{Perceived Evidence Grounding} & \textbf{Appropriateness} & \textbf{Reasoning Support} \\
\midrule

\multirow{3}{*}{Risk-Aligned}
& No-XAI
& 4.16 $\pm$ 0.93
& 4.42 $\pm$ 0.84
& 3.96 $\pm$ 0.85
& 4.13 $\pm$ 0.92
& 4.09 $\pm$ 0.82 \\

& XAI-Only
& 4.27 $\pm$ 0.80
& 4.27 $\pm$ 0.77
& 3.89 $\pm$ 0.99
& 4.14 $\pm$ 0.86
& 3.92 $\pm$ 0.89 \\

& XAI + Persona
& 4.07 $\pm$ 0.89
& 4.24 $\pm$ 0.69
& 3.74 $\pm$ 0.94
& 4.38 $\pm$ 0.62
& 4.05 $\pm$ 0.82 \\

\midrule

\multirow{3}{*}{Risk-Diluting}
& No-XAI
& 2.68 $\pm$ 1.29
& 3.24 $\pm$ 1.09
& 3.03 $\pm$ 1.12
& 3.16 $\pm$ 1.34
& 3.08 $\pm$ 1.21 \\

& XAI-Only
& 3.00 $\pm$ 1.41
& 3.50 $\pm$ 0.97
& 3.66 $\pm$ 1.02
& 3.22 $\pm$ 1.27
& 3.14 $\pm$ 1.05 \\

& XAI + Persona
& 2.71 $\pm$ 1.27
& 3.49 $\pm$ 1.05
& 3.22 $\pm$ 1.04
& 3.17 $\pm$ 1.12
& 2.90 $\pm$ 1.07 \\

\bottomrule
\end{tabular}
}
\end{table*}

\textbf{Human Evaluation Findings.}  Table~\ref{tab:human_results} summarizes human evaluation results across risk-aligned and risk-diluting conditions. Overall, human evaluators exhibit clear sensitivity to semantic degradation under risk-diluting conditions, assigning lower scores across most interpretation-oriented dimensions.

\indent \textbf{(a) Sensitivity to Semantic Misalignment.}  Under risk-aligned conditions, all explanation settings receive relatively high scores across Helpfulness, Clarity, Appropriateness, and Reasoning Support, generally within the $4.0$--$4.4$ range. This suggests that semantically aligned explanations are perceived as coherent, informative, and contextually appropriate regardless of explicit grounding configuration.  In contrast, risk-diluting conditions produce substantial degradation across interpretation-oriented dimensions. For example, risk-diluting \textit{No-XAI} explanations show clear declines in Helpfulness ($2.68 \pm 1.29$), Appropriateness ($3.16 \pm 1.34$), and Reasoning Support ($3.08 \pm 1.21$) relative to their risk-aligned counterparts. Similar degradation patterns appear across risk-diluting \textit{XAI-Only} and \textit{XAI + Persona} conditions.

These findings suggest that human evaluators remain broadly sensitive to semantically weakened or misleading explanations, particularly in dimensions associated with explanatory usefulness and contextual appropriateness.

\indent \textbf{(b) Grounding Illusions in Human Evaluation.}  
Despite overall degradation under risk-diluting conditions, Perceived Evidence Grounding remains comparatively resilient across grounded settings. Interpretation-oriented dimensions nevertheless continued to degrade under risk-diluting conditions, suggesting that human evaluators remained partially sensitive to semantic weakening despite comparatively resilient grounding perceptions. Notably, risk-diluting \textit{XAI-Only} explanations achieve the highest Perceived Evidence Grounding score ($3.66 \pm 1.02$) among risk-diluting conditions despite modest scores in Helpfulness, Appropriateness, and Reasoning Support. This pattern reflects a human-side \textit{grounding-illusion gap}, where explanations continue to appear grounded despite weakened interpretation. Explanations referencing detector-derived evidence may therefore appear trustworthy even when they fail to support the detector’s intended risk assessment. Semantically aligned XAI-grounded explanations also do not consistently receive higher Perceived Evidence Grounding scores than ungrounded explanations, suggesting that human grounding perception does not directly track attribution-based grounding or semantic faithfulness.


Although human evaluators exhibit the same qualitative grounding-illusion pattern as LLM judges, the overall magnitude is substantially smaller. Humans generally use a narrower scoring range and assign comparatively moderate scores even under risk-diluting conditions, whereas LLM-based judges frequently assign near-minimum scores under semantically misaligned settings.

These observations suggest that automated evaluators may apply more extreme penalties under semantic degradation, whereas human judgments remain comparatively more calibrated across dimensions. At the same time, the persistence of elevated Perceived Evidence Grounding scores under semantically weakened conditions indicates partial human susceptibility to grounded-looking but misleading explanations.

\indent \textbf{(c) Statistical Evidence for the Grounding-Illusion Framework.}  
Table~\ref{tab:planned_contrasts} reports selected paired comparisons from the human evaluation. Risk-diluting \textit{XAI-Only} explanations show significant degradation in Helpfulness and Appropriateness, whereas Perceived Evidence Grounding exhibits smaller and statistically non-significant degradation after Holm correction. Moreover, under risk-diluting \textit{XAI-Only} conditions, Perceived Evidence Grounding remains significantly higher than Reasoning Support, consistent with the grounding-illusion framework and proposed grounding-illusion gap.

\begin{table}[htbp]
\centering
\scriptsize
\caption{Selected paired comparisons from the human evaluation. RA and RD denote Risk-Aligned and Risk-Diluting conditions, respectively.}
\label{tab:planned_contrasts}
\setlength{\tabcolsep}{3pt}
\begin{tabular}{p{1.5cm}p{1.5cm}cccc}
\toprule
\textbf{Contrast} & \textbf{Metric} & \textbf{Mean Diff.} & \textbf{95\% CI} & \textbf{Holm $p$} & \textbf{$d_z$} \\
\midrule

\multirow{3}{*}{\makecell[l]{RA vs.\\ RD XAI-Only}}
& Helpfulness
& -1.41
& [-2.10, -0.72]
& .0067
& -0.91 \\

& Perceived Evidence Grounding
& -0.39
& [-0.94, 0.16]
& .638
& -0.31 \\

& Appropriateness
& -1.08
& [-1.58, -0.57]
& .0047
& -0.95 \\

\midrule

RD XAI-Only
& Perceived Evidence Grounding $-$ Reasoning Support
& 0.50
& [0.29, 0.71]
& $< .001$
& 0.96 \\

\bottomrule
\end{tabular}
\end{table}

Because descriptive table values are computed over participant-item ratings whereas inferential comparisons use participant-condition aggregates, reported mean differences do not necessarily match direct differences between marginal table means. Nevertheless, the paired comparisons support the grounding-illusion framework: perceived grounding remains comparatively resilient even when interpretation-oriented explanation quality declines.

\indent \textbf{(d) Persona-Adaptive Framing Effects.}  Persona-adaptive framing introduces comparatively modest variation in human evaluation outcomes. Across both risk-aligned and risk-diluting settings, the \textit{XAI + Persona} condition does not consistently outperform the corresponding \textit{XAI-Only} condition across interpretation-oriented metrics.

Under risk-diluting conditions, \textit{XAI + Persona} explanations frequently receive slightly lower Helpfulness and Reasoning Support scores than risk-diluting \textit{XAI-Only} explanations. This suggests that stylistic framing alone is insufficient to overcome semantic degradation when the underlying interpretation of detector-derived evidence becomes misleading.

From a trustworthy-AI perspective, these findings indicate that explanations may remain contextually appropriate and stylistically reassuring while still weakening the detector's semantic risk assessment. More broadly, the results suggest that human evaluators remain partially susceptible to grounded-looking but semantically weakened explanations under risk-diluting conditions.

\section{Discussions: Trustworthy Grounding and Semantic Alignment}
\label{sec:discussion}

This study reveals a consistent divergence between lexical grounding and semantic risk alignment in AI-assisted scam-risk explanations. Across both automated and human evaluations, explanations may continue to appear grounded by referencing detector-derived evidence even when their underlying interpretation becomes weakened or misleading.

\indent \textbf{(a) Evidence Citation Is Not Semantic Alignment.}  
The primary insight of this work is the empirical characterization of the \textit{grounding illusion}. Under risk-diluting conditions, explanations preserve grounded-looking evidence references while weakening the detector's semantic risk assessment.  Across both automated and human evaluations, risk-diluting grounded explanations retain comparatively elevated Perceived Evidence Grounding scores despite degradation in interpretation-oriented dimensions such as Helpfulness, Appropriateness, and Reasoning Support. These findings suggest that surface-level evidence citation alone is insufficient for trustworthy explanation evaluation.

\indent \textbf{(b) Perceived Grounding versus Attribution Faithfulness.}  Perceived Evidence Grounding does not necessarily correspond to attribution-based grounding or semantic faithfulness. Even risk-diluting \textit{No-XAI} explanations occasionally receive elevated grounding scores, suggesting that both humans and LLM judges may infer grounding from fluent message-specific references alone. At the same time, LLM-based judges exhibit larger grounding-illusion gaps under risk-diluting conditions, whereas human evaluators demonstrate smaller but still persistent grounding-illusion effects.

\begin{table}[htbp]
\centering
\footnotesize
\caption{Grounding-illusion gaps across evaluators under the Risk-Diluting XAI-Only condition.}
\label{tab:illusion_gap}
\setlength{\tabcolsep}{3pt}
\resizebox{\linewidth}{!}{
\begin{tabular}{lccc}
\toprule
\textbf{Evaluator} & \textbf{Perceived Evidence Grounding} & \textbf{Semantic Avg.} & \textbf{Illusion Gap} \\
\midrule
Human & 3.66 & 3.12 & 0.54 \\
GPT-5.4 & 3.44 & 1.52 & 1.92 \\
Gemini-2.5-Pro & 4.50 & 1.50 & 3.00 \\
\bottomrule
\end{tabular}
}
\end{table}

\indent \textbf{(c) Evaluation Reliability and Trust Calibration.}  
The grounding-illusion gap highlights an important reliability challenge for human-facing AI systems. Automated evaluators may remain overly sensitive to grounded-looking evidence references even when semantic interpretation deteriorates, potentially overestimating explanation trustworthiness.

For security-critical systems, lexical grounding alone should not be treated as sufficient evidence of trustworthy explanation behavior. Grounded-looking but semantically misleading explanations may distort user trust calibration by encouraging over-reliance on AI-assisted security assessments.

Accordingly, future explanation systems may require auditing not only for evidence citation fidelity but also for semantic risk preservation and trust calibration. More broadly, our findings suggest that lexical grounding, semantic alignment, and attribution faithfulness should be evaluated as distinct properties rather than interchangeable quality proxies.

These concerns may be especially consequential for digitally vulnerable populations, including users with limited cybersecurity literacy or familiarity with AI-assisted decision systems, who may rely more heavily on AI-generated explanations during security-related decision making.

\section{Conclusions and Future Work}
\label{sec:conclusion}

This work provides evidence that evidence citation alone is insufficient for semantic risk alignment in AI-assisted scam-risk explanations. Across both automated and human evaluations, explanations may preserve grounded-looking references to detector-derived evidence while weakening the intended risk interpretation, revealing grounding-illusion effects in which Perceived Evidence Grounding remains resilient despite semantic degradation. These findings highlight the importance of evaluating not only whether explanations reference evidence, but also whether they interpret it consistently with the intended security assessment. More broadly, trustworthy explanation systems should distinguish lexical grounding, semantic alignment, and attribution faithfulness when evaluating explanation reliability.

Several limitations motivate future work. The current human evaluation remains relatively limited and may reflect demographic or security-literacy biases associated with university-affiliated recruitment. In addition, risk-diluting explanations were generated under controlled conditions rather than deployed systems. Future work should therefore evaluate grounding illusions across broader populations, real-world prompt injection and conversational-assistance scenarios, detector uncertainty conditions, and downstream outcomes such as trust calibration and security decision making. Future evaluation frameworks should further support joint assessment of evidence grounding and semantic risk preservation.

\clearpage
\section*{Ethical Considerations}
\label{app:ethics}

This study involves human evaluation and publicly available scam-risk datasets spanning email, SMS, and social media. All human-study responses were collected anonymously, and participation was voluntary. Participants provided informed consent prior to participation, and the study protocol was approved by the institutional IRB.

The scam-risk corpora used in this study were obtained from publicly available research datasets. Example messages included in the paper were manually reviewed to avoid unnecessary disclosure of sensitive or identifying information.

Because VEXA includes prompts for generating risk-diluting explanations, releasing full adversarial prompts or raw scam messages could create dual-use risks. Accordingly, we plan to release only sanitized examples, aggregate evaluation data, and controlled prompt templates that prevent direct reuse for scam facilitation. Public artifacts will remove URLs, phone numbers, account identifiers, and other sensitive information.

More broadly, this work aims to improve the robustness of AI-assisted security explanations by highlighting the limitations of relying solely on evidence citation without verifying semantic alignment and risk interpretation.

We avoid interpreting style-adaptive framing conditions as normative or clinical claims about individual users or psychological vulnerability. These stylized framings are used solely to examine how explanation style influences evaluation robustness under controlled semantic conditions.  We do not interpret these stylized framings as direct indicators of real-world psychological vulnerability or user personality.

\bibliography{ref.bib}

\appendix

\end{document}